\documentstyle[11pt,paspconf,epsf]{article}
\def\lesssim{\mathrel{\hbox{\rlap{\hbox{\lower4pt\hbox{$\sim$}}}\hbox{$<$}}}}
\def\gtrsim{\mathrel{\hbox{\rlap{\hbox{\lower4pt\hbox{$\sim$}}}\hbox{$>$}}}}
\def\sq{\hbox{\rlap{$\sqcap$}$\sqcup$}}
\newcommand{\ropt}{R$_{25}$ }
\newcommand{\ha}{H$\alpha$ }

\begin{document}

\title{The Extreme Outer Regions of Disk Galaxies: Star Formation and
Metal Abundances}
\author{Annette Ferguson}
\affil{Institute of Astronomy, Univ. of Cambridge, Cambridge UK CB3 0HA}
\author{Rosemary Wyse}
\affil{Dept. of Physics \& Astronomy, JHU, Baltimore, MD, USA 21218}
\author{Jay Gallagher}
\affil{Dept. of Astronomy, University of Wisconsin, Madison, WI, USA 53706}

\begin{abstract}
The extreme outer regions of disk galaxies, lying at or beyond the
classical optical radius defined by R$_{25}$, present an opportunity to
study star formation and chemical evolution under unique physical
conditions, possibly reminscent of those which existed during the early
stages of disk evolution.  We present here some of the first results
from a large study to measure star formation rates and metallicities in
the extreme outer limits of a sample of nearby spiral galaxies.
Despite their low gas column densities, massive star formation is often
observed in these outer parts,  but at an azimuthally--averaged rate
much lower than that seen in the inner disk.  Gas-phase O/H abundances
of roughly 10\% solar characterize the gas at 1.5--2~R$_{25}$.  The
implications of our results for star formation $`$laws' and models of
disk evolution are discussed.
\end{abstract}

\keywords{disk galaxies; star formation; chemical abundances}

\section{Introduction}

Distinguishing between competing models of disk galaxy formation and
evolution requires observational constraints on the radial variations
of the present star formation rate, the star formation history, and the
gas-phase chemical abundance.  Unfortunately, most observational
studies  to date have focused only on the bright, easily-observed inner
regions of galactic disks, lying at or within the classical optical
radius, R$_{25}$ (defined by the B-band 25th magnitude isophote).  It
is well known, however,  that disk galaxies have HI  disks which extend
to typically $\gtrsim$ 1.5--2~\ropt, and in some rare cases to
$\gtrsim$~3~R$_{25}$.  Knowledge of the star formation rates and
metallicities in these optically-faint, extreme outer reaches of disks
is of particular importance for a variety of reasons.  First of all,
the predictions of various star formation laws and chemical evolution
models often diverge most strongly in the outer parts of galaxies (eg.
Prantzos \& Aubert 1995, Tosi 1996), hence observational constraints
are needed as far out in the disk as possible.  Furthermore,  the outer
regions of disk galaxies provide a unique opportunity to study star
formation  and chemical evolution under the extreme physical conditions
of low gas surface density (yet high gas fraction), low metallicity and
long dynamical times; similar conditions are also inferred for many
high-redshift  damped Lyman-$\alpha$ systems (eg. Pettini et al 1997),
and low surface brightness galaxies (Pickering et al 1997).

We have carried out a large observational project to study the extreme
outer disks of a sample of nearby spiral galaxies.  Results are
presented here for three galaxies (NGC~628, NGC~1058 and NGC~6946)
studied so far which exhibit recent massive star formation at
particularly large radii (and which perhaps not surprisingly  have
unusually large HI-to-optical sizes).  A full discussion of these
results is provided in Ferguson et al (1998a, 1998b).  Analysis of our
larger sample ($\sim$15 galaxies) is currently ongoing, and results
will be presented at a later date (Ferguson et al, in prep).

\section{Observations}

Deep wide-field H$\alpha$ images were obtained to map the distribution
of recent star formation, using the KPNO 0.9~m and the Lowell 1.8~m
telescopes (deep BVR images were also obtained to study the extent and
morphology of the underlying stellar disks, but these data will not be
discussed here).  Radial  H$\alpha$ surface brightness profiles were
constructed via elliptical aperture photometry on the H$\alpha$
images.

Long-slit spectroscopy in the range 3700--7000\AA\ was carried out for
a small sample ($\sim$~10) of our newly-discovered outer disk HII
regions, in order to obtain metallicities.  Oxygen and nitrogen
abundances were derived via well-established $`$semi-empirical'
methods, based on the inter-relationship between metallicity and the
intensities  of the strong lines, [OII] $\lambda$3727 and [OIII]
$\lambda\lambda$4959,5007, via the parameter $R_{23}$. We adopted  the
particular calibrations proposed by McGaugh (1991) and Thurston et al
(1996) to derive O/H and N/O respectively.

In Figure 1, we show a continuum--subtracted H$\alpha$ images of one
of the galaxies, NGC~6946, with the HII regions for which
we obtained spectra identified.

\begin{figure}[h]
\plotfiddle{ferg.fig1c.ps}{6cm}{270}{50}{50}{-200}{250}
\vspace*{1cm}
\caption{An H$\alpha$ continuum-subtracted image of NGC~6946. The marked HII
regions indicate those for which metallicities have been obtained.
R$_{25}$ is marked in each case by the large dashed circle.}
\end{figure}

\section{Star Formation Beyond the Optical Edge of Disk Galaxies}

\subsection{Morphology}

Our deep images reveal the discovery of recent massive star formation
-- HII regions -- out to the extent of our imagery
($\gtrsim$2~R$_{25}$) in all three galaxies.  The inner and outer disk
HII regions appear strikingly different, in that  star formation in the
outer disk occurs in  smaller, fainter and more isolated HII regions.
The brightest outer disk HII regions detected here have diameters of
150--500 pc and \ha luminosities  of only
1--80~$\times$~10$^{37}$~erg~s$^{-1}$ (for reference, the Orion nebula
has L$_{H\alpha}$~$\sim$~10$^{37}$~erg~s$^{-1}$).  These luminosities
imply  enclosed ionizing populations of 0.2--20 equivalent O5V stars.
Establishing whether the populations of inner and outer disk HII
regions are actually intrinsically different, perhaps reflecting
different modes of star formation, or if they only appear that way due
to poor statistical sampling  of the HII region luminosity function in
the outer disk (ie. fewer HII regions, hence fewer luminous ones) is an
important issue that we will address in the future.

The outer disk star formation appears remarkably organized, with the
HII regions delineating narrow spiral arms. The pattern is present in
all cases, but more obvious in NGC~628 and NGC~6946. Our deep
broad-band images reveal the existence of faint (B~$\sim$~26--28
mag/$\sq$\arcsec) stellar arms in all three galaxies, associated with
these HII arms, and inspection of published HI maps also reveals similar
structures in the underlying neutral gas (eg. Shostak \& van der Kruit
1984; Dickey et al 1990; Kamphuis 1993).  The relationship between the
inner and outer spiral structure remains unclear, as indeed is the
dynamics underlying the outer arms. The lack of obvious companions to
any of these galaxies makes the tidal hypothesis for spiral arm
formation unlikely; future observations in the near-IR will help to
distinguish between alternative theories for the outer spiral patterns,
such as long--lived density waves, or transient shearing
perturbations.

\subsection{Star Formation Rates}

The  radial variation of the massive star formation rate per unit area
across the disk is traced by the azimuthally--averaged H$\alpha$
surface brightness ($\Sigma_{H\alpha}$) distribution.   Since most
models for star formation invoke dependences on some form of the gas
density, it is informative to plot $\Sigma_{H\alpha}$ against various
components of the interstellar gas (Figure 2).  The different
components of the gas are seen to correlate in very different ways with
the star formation rate.  While a roughly linear correlation is seen
between $\Sigma_{H\alpha}$  and $\Sigma_{CO}$ (except for NGC~1058,
where the relation is somewhat steeper), a very complicated non-linear
behaviour is seen between $\Sigma_{H\alpha}$  and  $\Sigma_{HI}$.
Abrupt steepenings are seen at low total gas column densities in all
cases, where they occur at azimuthally-averaged total gas surface
densities of 5--10~M$_{\sun}$/pc$^2$ or
6--10~$\times$~10$^{20}$~cm$^{-2}$.  Inspection of Figure 2 indicates
that the location of the steepenings is approximately coincident with
the radius where the disk undergoes the transition from being dominated
by (warm) molecular to atomic gas, suggesting that a high covering
factor of the molecular phase  might be a requisite for significant
star formation (eg. Elmegreen \& Parravano 1994).

\begin{figure}[h]
\plottwo{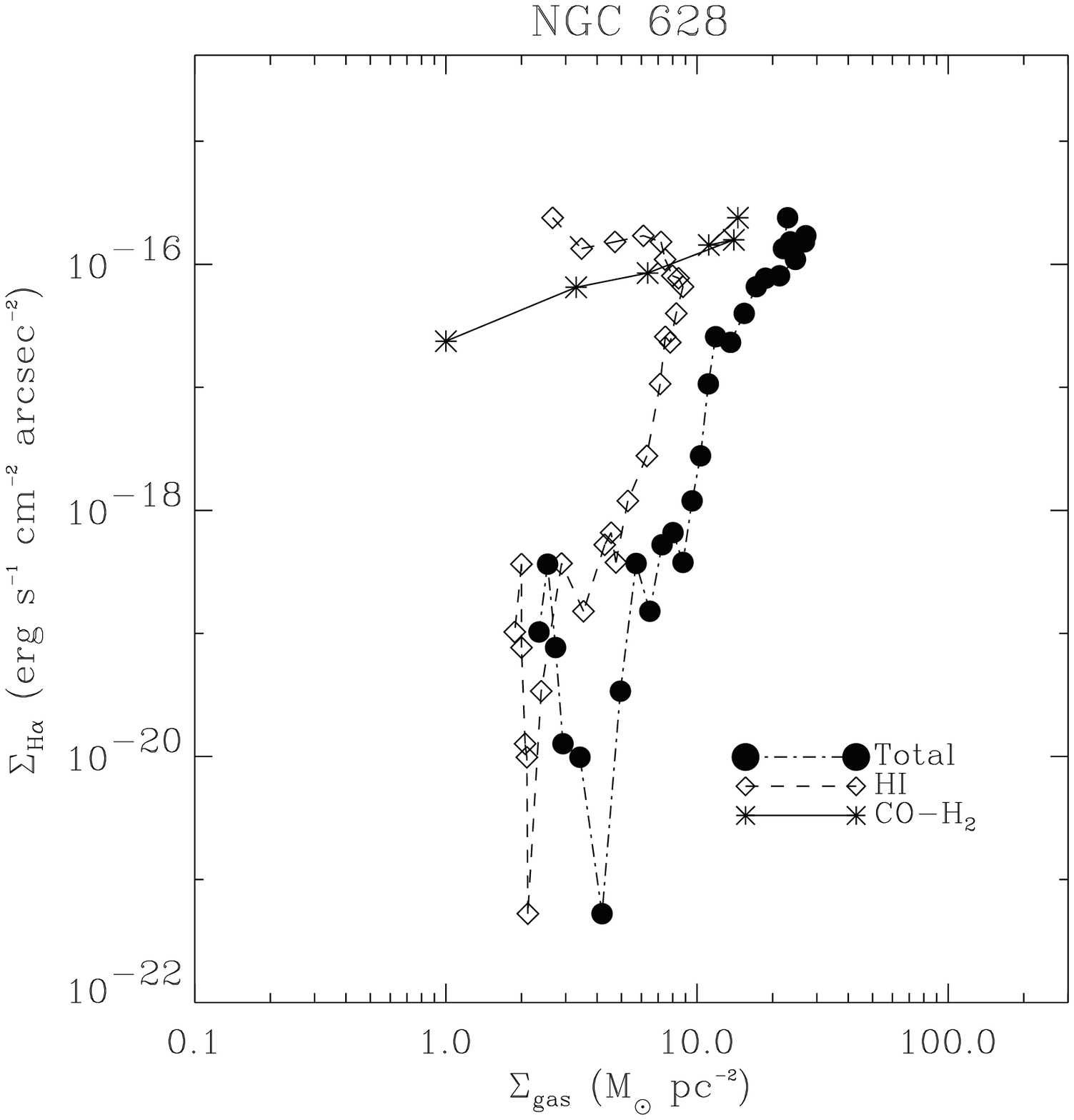}{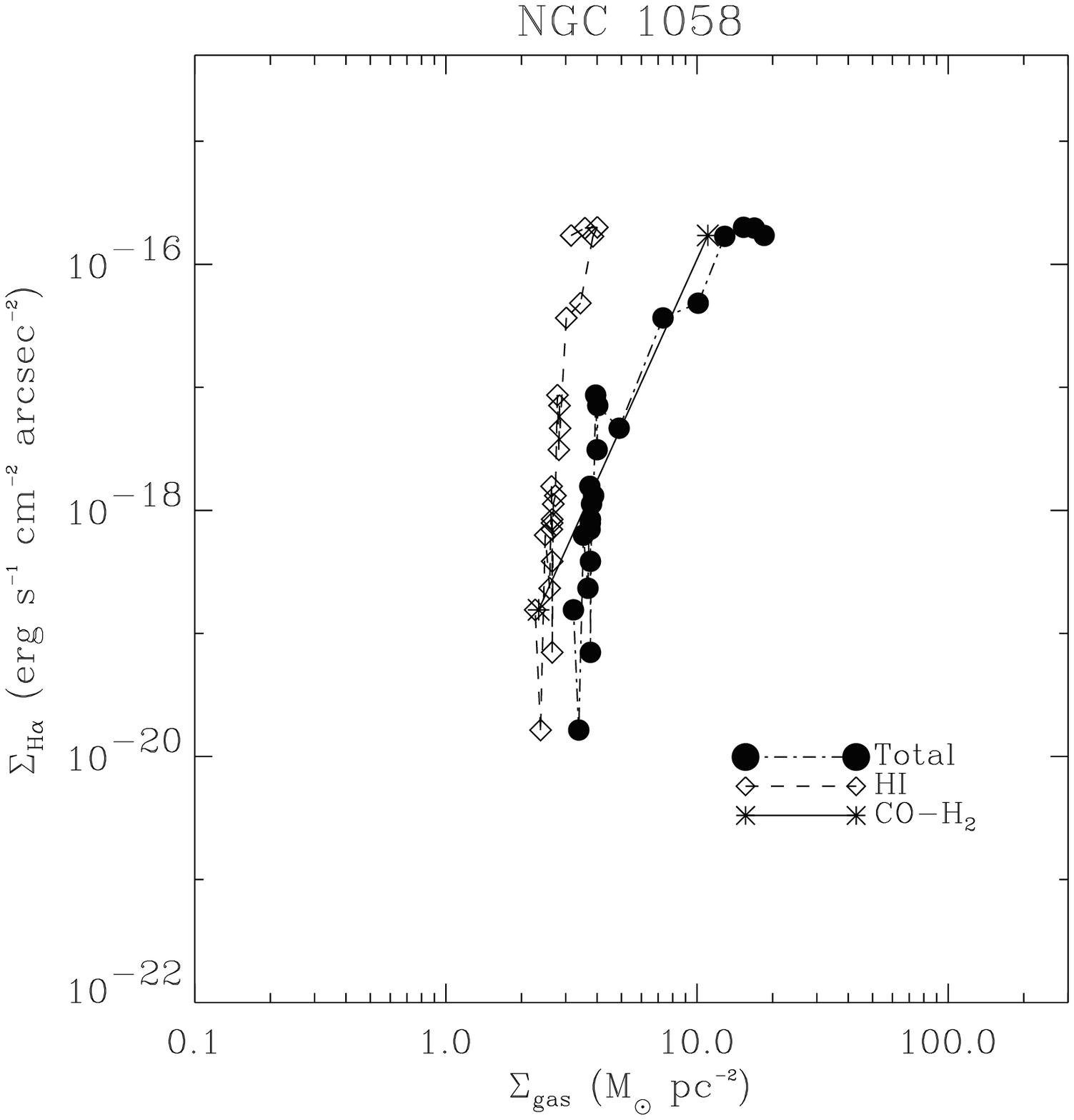}
\vspace*{0.5cm}
\plotfiddle{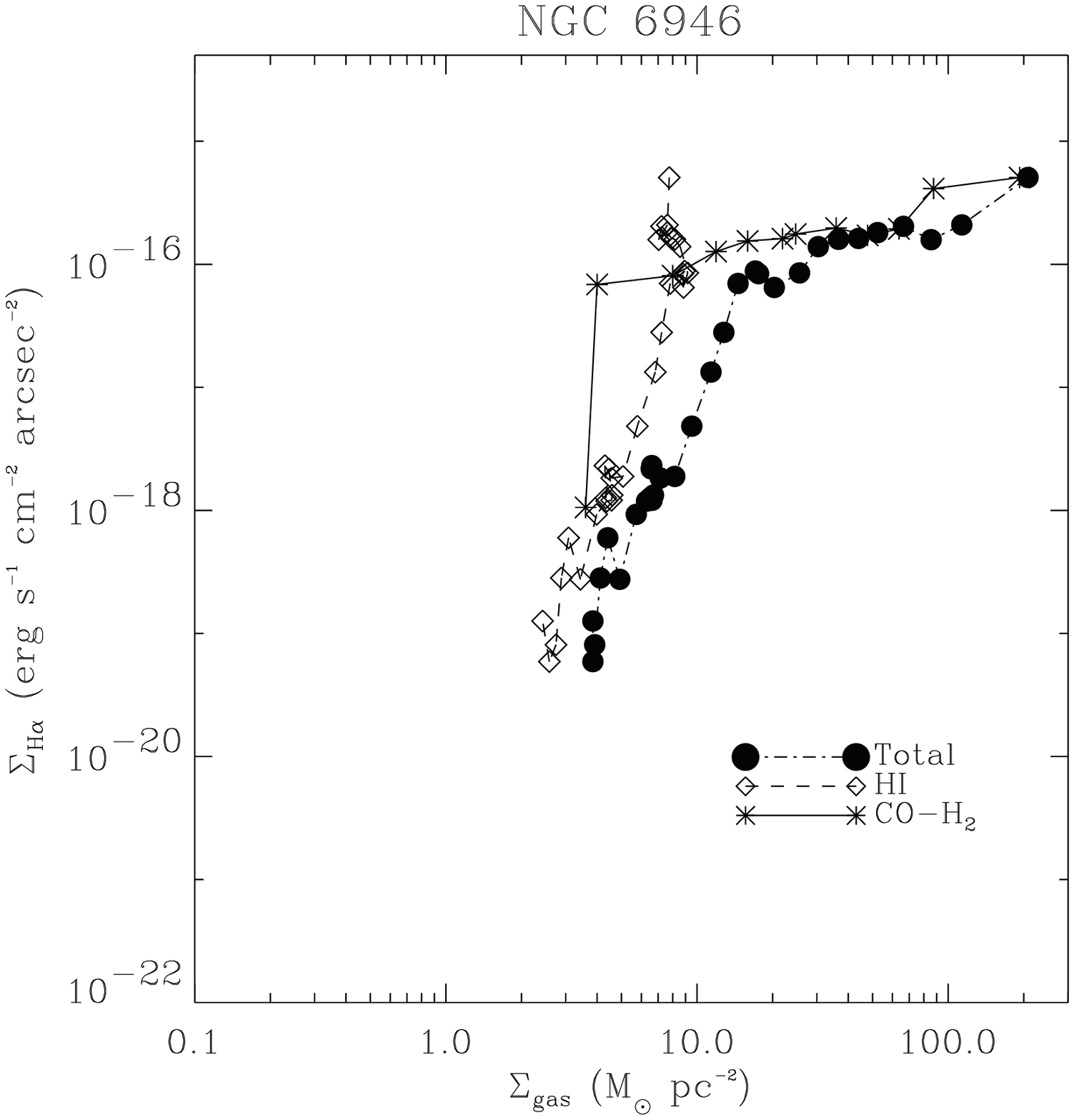}{4cm}{0}{33}{33}{-130}{-60}
\vspace*{0.6cm}
\caption{ The variation of the
deprojected, azimuthally-averaged  H$\alpha$ surface brightness,
$\Sigma_{H\alpha}$, versus various components of gas surface density, $\Sigma_{gas}$
(both quantities have been determined as a function of galactocentric
radius in a series of elliptical annuli). The total gas surface density is
the sum of the $\Sigma_{CO}$ and $\Sigma_{HI}$, corrected for heavy elements. }
\end{figure}

The existence of abrupt declines in $\Sigma_{H\alpha}$ at low gas
surface densities make it impossible to describe the observations with
a single-component Schmidt law, with a dependence on total gas surface
density  alone.  These  steepenings are due  almost entirely  to sharp
declines in the {\it covering factor} of star formation however, and
not to changes in the rate at which stars form locally (see Ferguson et
al 1998b).  In other words, whenever star formation occurs, it occurs
with roughly the same local intensity in both the inner and outer
disks; it is the processes which lead to star formation, not the star
formation itself, that are apparently much less efficient in the outer
disk than in the inner disk.

\section{What Drives Star Formation at Low Gas Surface Densities? }

The instability driving star formation may well be gravitational; in this 
case the Toomre-Q criterion (Toomre 1964) 
yields a critical gas surface density above which one expects local 
instability to axisymmetric perturbations.  For an
infinitely thin, one component isothermal gas disk, the critical gas
surface density above which self-gravity overcomes shear and pressure is 
given by 
\begin{equation}
\Sigma_{crit}=\frac{{\alpha\sigma}{\kappa}}{\pi{G}} \end{equation}
where  $\alpha$ is a constant of order unity, $\sigma$ is the 
velocity dispersion of the gas  and $\kappa$ is the epicyclic
frequency\footnote{The epicyclic frequency is defined as
$\kappa= \sqrt{2}~{\frac{V}{R}}\sqrt{1+{\frac{R}{V}}~{\frac{dV}{dR}}}$.}.
Following Kennicutt (1989), we adopt  $\alpha=$0.67.

We have used published gas data (rotation curves and velocity
dispersions) to calculate the radial variation of $\Sigma_{crit}$  for
each galaxy.  Both NGC628 and NGC1058 have low inclinations, which
means that the amplitudes and shapes of the rotation curves are
somewhat uncertain (but, on the other hand, the radial variation of the
velocity dispersions are known with good accuracy). For these two
galaxies, we have normalised the rotation curves using the Shostak
(1978) relation between M$_B$ and the maximum rotation velocity,  and
assumed that the curves remain flat in the outer regions.   Two
estimates of the critical gas density were made, one assuming a
constant gas velocity dispersion (6~km~s$^{-1}$ for consistency with
Kennicutt 1989) and the other including the radial variation determined
directly from the HI observations.

\begin{figure}[h]
\plotfiddle{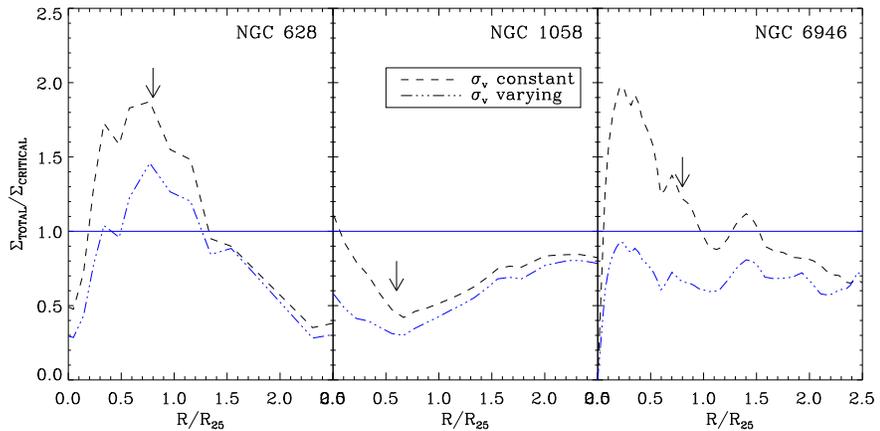}{4cm}{90}{50}{50}{180}{-60}
\vspace*{1.5cm}
\caption{The radial variation of the ratio of the
azimuthally--averaged (observed) gas surface density to the critical
surface density for gravitational instability, calculated using both a
constant velocity dispersion (dashed line) and a radially-varying one
(dashed-dotted line).  The solid horizontal line indicates the value of
the ratio above which instability is expected. }
\end{figure}

The observed gas surface densities typically lie within a factor of two of the
estimated critical densities at all radii, with the agreement often
being even better when a radially-varying velocity dispersion is used
(eg. NGC~6946, see Figure 3). This general agreement is very
encouraging, in view of that fact that many of the  input data -- eg.
rotation curves, gas surface densities, value of $\alpha$ etc -- used
to estimate $\Sigma_{crit}$  have significant  uncertainties.  It
therefore appears that the outer disks of these galaxies lie close
enough  to the Q-stability limit so that processes such as swing
amplification can operate and trigger star formation locally.
Realistic uncertainties in the quantities used in this derivation are
not likely to change the value of the stability parameter
($\Sigma_{crit}$/$\Sigma_{crit}$) by more than a factor of two at any
given radius, and hence should not affect our principal conclusion that
the gas disks are marginally unstable over a  large radial zone.

On the other hand, the abrupt decreases in star formation rate at low
gas surface density appear uncorrelated with changes in the
gravitational stability of the disk (see Figure 3); that is, the
locations where $\Sigma_{H\alpha}$ plummets (indicated by the small
arrows in Figure 3) are no more or no less stable than any other
location in the disk.  Further, while the stability parameter changes
by only a factor of a few at most across each disk, the star formation
rate per unit area typically changes by a factor of 10$^2$--10$^3$!  
The sizes of the star formation regions are also difficult
to reconcile with the predictions of the gravitational instability
model, the theoretically most unstable length being much larger (ie. a
few kiloparsecs).  Thus, while local gravitational instability can
probably account for the existence of star formation at large radii in
these three galaxies, it fails to explain either the rates of star
formation or the  scales of star formation across the disks.

What alternative explanations are there for the reduced star formation
rates seen in the outer disk?  It may be that star formation occurs
only when  the local gas column exceeds some fixed threshold value
(eg.  Skillman 1987), and perhaps reaching that threshold becomes more
difficult at larger radii. Such a local threshold could  be related to
a critical column density of dust necessary to shield molecular gas
from UV radiation.  Indeed, Elmegreen \& Parravano (1994) argue that
the low thermal pressures make it increasingly difficult to sustain a
cool molecular phase beyond $\sim$3 disk scale-lengths. Comparison of
our H$\alpha$ images to HI maps shows star formation down to local HI
columns of a few times 10$^{20}$~cm$^{-2}$, but the HI maps are for the
most part low resolution (few kiloparsecs) and better data are needed
to test this idea.    Yet another explanation for the reduced rates
could be an intrinsic correlation between azimuthally--averaged star
formation rate and gas {\it volume\/} density, combined with a vertical
flaring of the gas disk, such that the transformation between gas
surface density and volume density varies with galactocentric radius
(see also Madore et al 1974). It is well known that gaseous disks
exhibit an increase of scaleheight at large radius (eg.  Merrifield
1992) whereas the young stars are confined to a thin plane.  The
combination of an increase in gas scale height with radius with the
slow decline of gas surface density could conspire to produce a rapid
decline in areal star formation rate between the inner and the outer
disks, as is observed.  We will investigate the success of these
models in more detail in future work.

\section{Chemical Abundances at Large Radii}

Figure  4 presents the derived O/H abundances for the
three galaxies as a function of the deprojected
galactocentric radius, normalised to R$_{25}$. 
 The outermost abundances in all three galaxies are $\sim$~10-15\%
solar, measured at radii in the range 1.5--2~R$_{25}$.   Although these
represent some of the lowest abundances ever measured in spiral disks, they
indicate that the outermost gas is far from being pristine.
 
\begin{figure}[h]
\plottwo{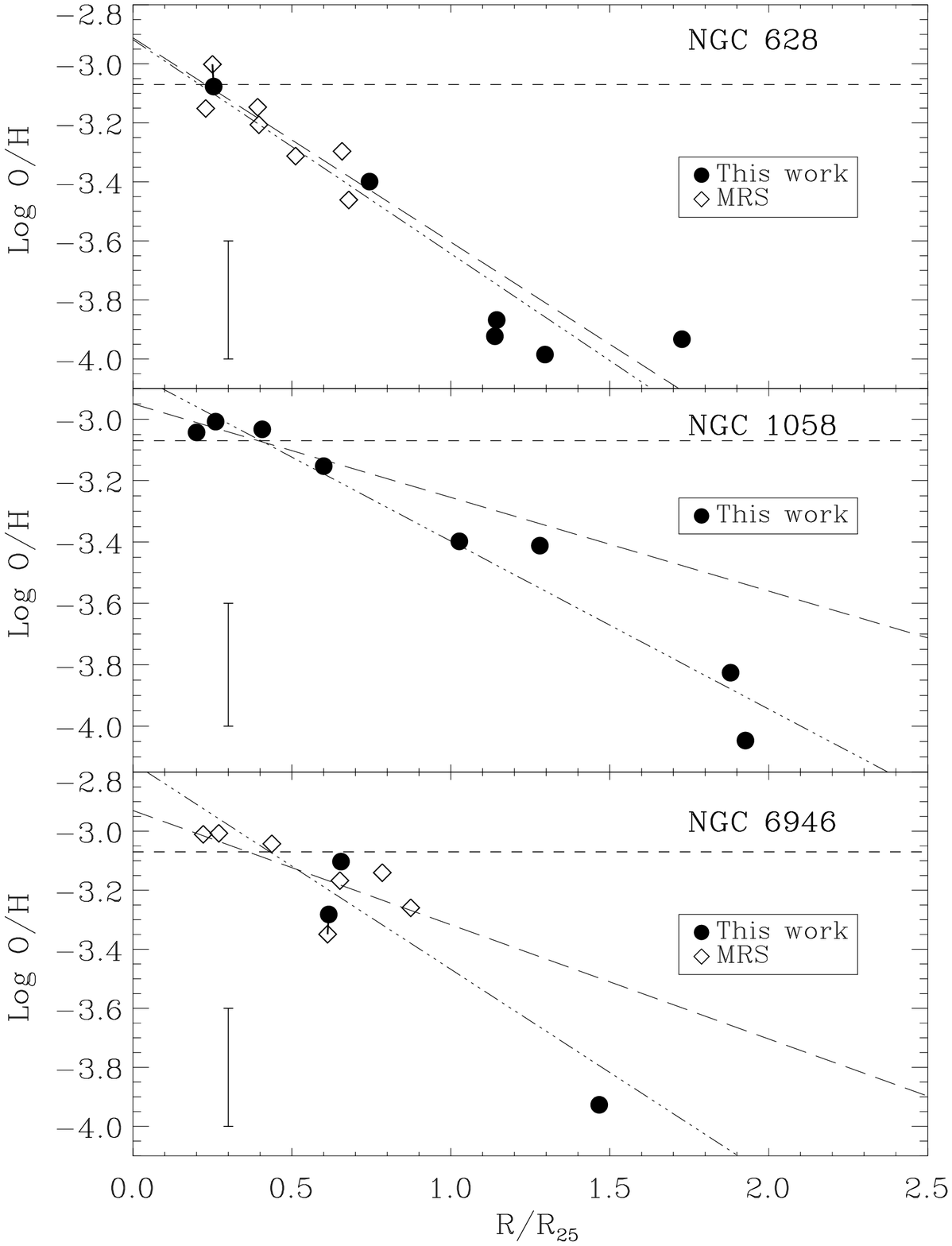}{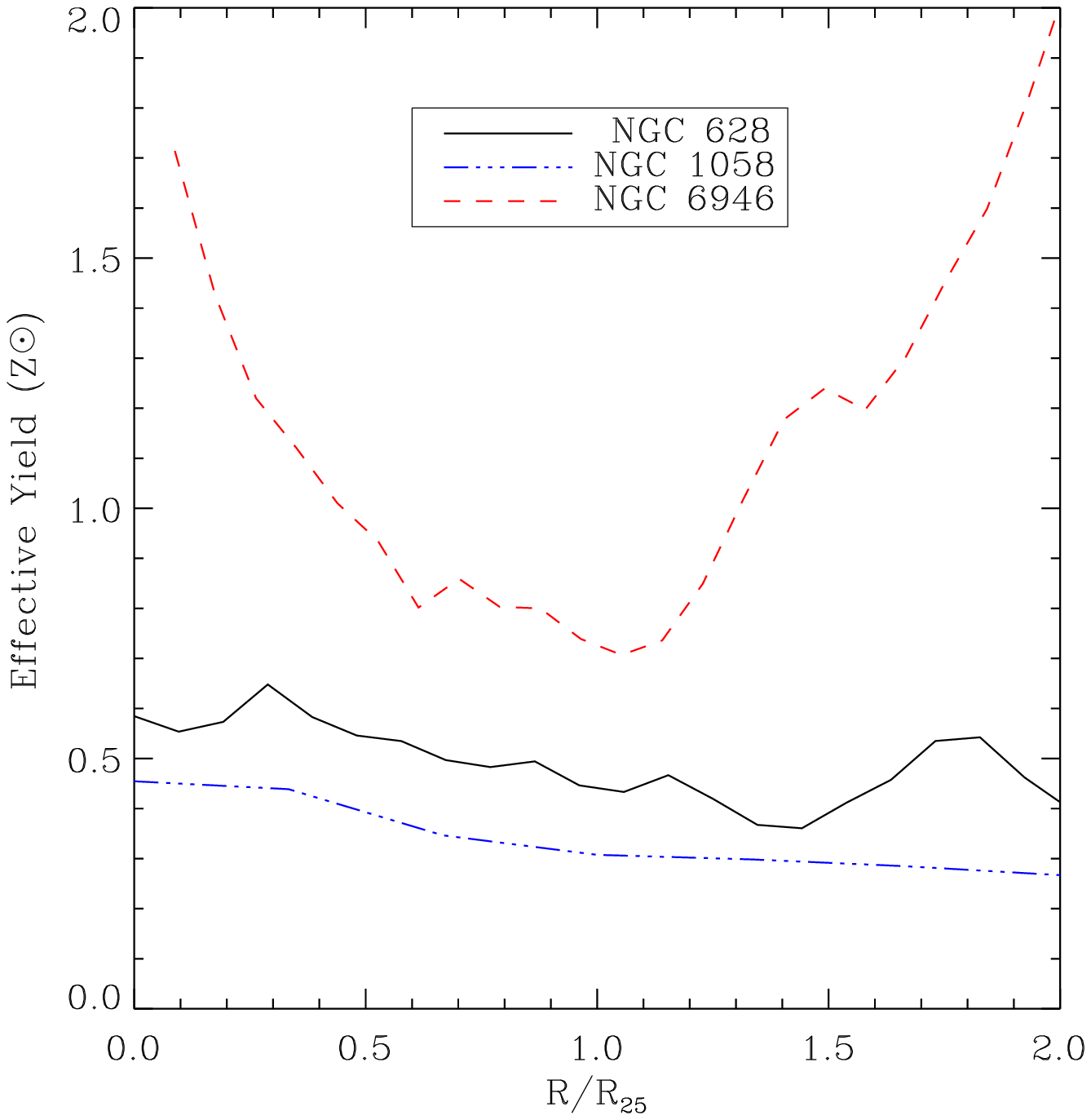}
\caption{(Left) Radial variation of O/H, expressed in terms of
the optical radius.  The solar value is indicated by the horizontal
dashed line. The dashed-dotted lines indicate fits to all the abundance
measurements, whereas the long-dashed lines indicate fits to only
the inner disk points.  (Right) The radial variation of the effective yield,
as derived for the closed box model. }
\end{figure}

We have compared the O/H gradients derived from fitting the entire set
of datapoints in each galaxy and from fitting only those points lying
within the optical disk (see Figure 4). Clearly, the outer disk
abundances  play a crucial role in defining the abundance gradient
across the disk, leading to  significantly steeper gradients  for
NGC~1058 and NGC~6946.  Within the limits of the current dataset, it
appears that the radial abundance gradients can be adequately described
by single log-linear relationships.  This result would appear to imply
a continuity in the star formation (ie.  metal production) process from
inner to extreme outer disk, whereas our direct observations of the
present star formation rate indicate that this is not the case.  Gas
flows and/or significant pre-enrichment of the disk gas may be required
to reconcile these observations.

To assess the importance that gas flows might have had in the evolution
of outer galactic disks, we have compared our data with the predictions
of the simple $`$closed box' model (no inflow or outflow from each
radial zone).   The closed box model can be represented by a simple
relation $Z=-p~ln~\mu$ where $p$ is the yield of the element in question
and $\mu$ is the gas fraction, defined as baryonic mass in gas to the
total baryonic mass (stars + gas).  We have used our deep B-band
surface photometry and published HI and CO maps to calculate the gas
fractions.   Figure 4 (right panel) shows the radial variation of the
effective yield required to reconcile our observations with the closed
box model. Interestingly, both NGC~628 and NGC~1058 are consistent with
relatively constant effective yields across their disks, with values
similar to that found in the solar neighbourhood (0.5~Z$_{\odot}$; Wyse
\& Gilmore 1995).  In these cases, it would therefore appear that the
role of gas flows in the evolution of the  disk is similar to that for
the solar neighbourhood.  On the other hand, NGC~6946 exhibits gas
fractions that are too high for the observed metallicity at both small
and large radius.

\section{Conclusions}

The first results from our ongoing study of extreme outer galactic
disks indicate that, in at least some galaxies, these regions are the
sites of ongoing massive star formation and have metallicities which
are low, although far from pristine.  Observations of star formation in
these low gas surface density parts are of particular importance for
testing current ideas about the processes which govern large scale
formation in galaxies.  Our analysis suggests that considerations of
gravitational instability alone cannot explain why the rate of star
formation drops so abruptly beyond the optical edges of disk galaxies.

The low metallicities in these regions are similar to those measured in
some high-redshift damped Lyman-$\alpha$ systems (eg. Pettini et al
1997) and suggest that  outer disks are relatively unevolved at the
present epoch.  Perhaps star formation has only recently began in these
parts, or perhaps the extreme outer disk has been forming stars for a
significant period of time, but at a such low rate that little
evolution has had the chance to take place.   Distinguishing between
these two alternatives requires establishing the mean age of the bulk
of the outer disk stars, and we are currently investigating this issue via
deep HST photometry of resolved stars in the outer parts of galaxies.

\end{document}